\documentclass[showpacs,prb,twocolumn]{revtex4}
\usepackage{textcomp}
\usepackage{amsmath}
\usepackage{amssymb}
\usepackage{graphicx}

\begin{document}

\title{Quantum non-demolition measurements of a qubit coupled to a
  harmonic oscillator}

\author{Luca Chirolli}
\email{luca.chirolli@uni-konstanz.de}
\author{Guido Burkard}
\email{guido.burkard@uni-konstanz.de}
\affiliation{RWTH Aachen University, D-52056 Aachen, Germany\\
Department of Physics, University of Konstanz, D-78457 Konstanz, Germany}

\begin{abstract}
We theoretically describe the weak measurement of a two-level system 
(qubit) and quantify the degree to which such a qubit measurement has 
a quantum non-demolition (QND) character. The qubit is coupled to a 
harmonic oscillator which undergoes a projective measurement. 
Information on the qubit state is extracted from the oscillator measurement 
outcomes, and the QND character of the measurement is inferred from 
the result of subsequent measurements of the oscillator.  
We use the positive operator valued measure (POVM) formalism to describe 
the qubit measurement. Two mechanisms lead to deviations from a perfect QND measurement: (i) the quantum fluctuations of the oscillator, and (ii) quantum tunneling between the qubit states $|0\rangle$ and $|1\rangle$ during measurements.
Our theory can be applied to QND measurements performed on superconducting 
qubits coupled to a circuit oscillator. 
\end{abstract}


\maketitle

\begin{figure}[t]
 \begin{center}
  \includegraphics[width=8.3cm]{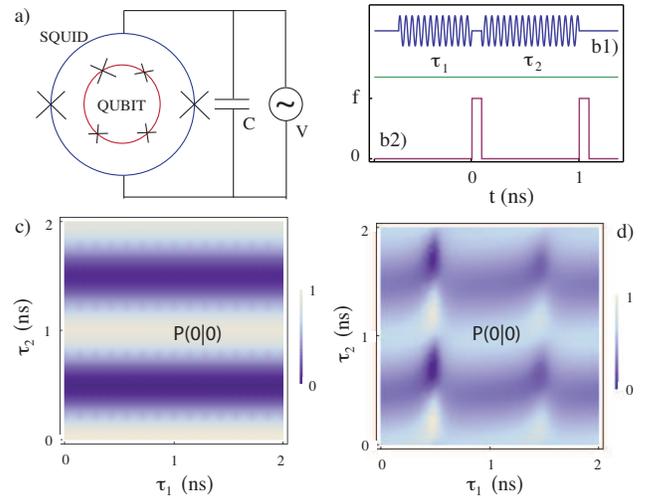}
    \caption{(Color online) a) Schematics of the 4-Josephson junction superconducting 			flux qubit surrounded by a SQUID. b) Measurement scheme: b1) two short 			pulses at frequency $\sqrt{\epsilon^2+\Delta^2}$,  before and between two  			measurements prepare the qubit in a generic state. Here, $\epsilon$ and 
    		$\Delta$ represent the energy difference and the tunneling amplitude 
		between the two qubit states. b2) Two pulses of amplitude $f$ and duration 
		$\tau_1=\tau_2=0.1~{\rm ns}$ drive the harmonic oscillator to a qubit-dependent 		state. c) Perfect QND: conditional probability $P(0|0)$ for $\Delta=0$ to detect 			the qubit in the state "0" vs driving time $\tau_1$ and $\tau_2$, at Rabi 				frequency of $1~{\rm GHz}$. 
		The oscillator driving amplitude is chosen to be 
		$f/2\pi=50~{\rm GHz}$ and the damping rate $\kappa/2\pi=1~{\rm GHz}$.   d) 
		Conditional probability $P(0|0)$ for $\Delta/\epsilon=0.1$, 
		$f/2\pi=20~{\rm GHz}$,  $\kappa/2\pi=1.5~{\rm GHz}$. A phenomenological 			qubit relaxation time $T_1=10~{\rm ns}$ is assumed. 
		\label{Fig1}}
 \end{center}
\end{figure}

\section{Introduction}

The possibility to perform repeated quantum measurements on a system 
with the least possible disturbance was first envisioned  in the context of 
measuring gravitational waves\cite{Braginsky-Khalili}. In quantum optics 
the optical Kerr effect provided an early playground for studying for QND 
measurements\cite{Milburn83,ImotoPRA85,BachorPRA88}, that were 
extended to the framework of cavity quantum electrodynamics (cavity-QED)
and mesoscopic mechanical oscillators. 
\cite{Fortunato99,ClerkPRB03,SantamorePRB04,DotsenkoPRA09,IrishPRB03}

The application of such a scheme to quantum information has stimulated 
great interest, in particular in the field of quantum computation, where  
fast and efficient readout is necessary, and error correction plays an
important role\cite{Nielsen}. 
Schemes for qubit QND measurement have been  theoretically proposed 
and experimentally realized with a superconducting qubit coupled to harmonic 
oscillators, either represented by external tank $LC$-circuit
\cite{IlichevPRL03,Chiorescu04,GrajcarPRB04,Bertet05,SillanpaaPRL05,DutyPRL05,
KatzScience06}, 
or by superconducting resonator that behaves as a one 
mode quantum harmonic oscillator in circuit-QED.  
\cite{Wallraff04,BlaisPRA04,SchusterPRL05,WallraffPRL05,
GambettaPRA06,GambettaPRA07,BoissonneaultPRA09}
A measurement scheme based on the Josephson 
bifurcation amplifier (JBA) \cite{SiddiqiPRL04,SiddiqiPRB06} 
has been adopted with the aim to perform QND measurements of 
superconducting qubit \cite{Lupascu07,BoulantPRB07}. In these 
experiments a deviation of $\sim 10\%$ from perfect QND behavior 
has been found.

Motivated by those recent experimental achievements we analyze a  
measurement technique based on the coupling of the qubit to a driven 
harmonic oscillator. A quadrature of the harmonic oscillator is addressed 
via a projective measurement. The qubit that is coupled to the oscillator 
affects the outcomes of the measurement of the oscillator and information 
on the qubit state can be extracted from the results of the projective 
measurement of the oscillator. We aim to shed some light on the 
possibilities to perform qubit QND 
measurements with such a setup, and to understand whether deviations 
from the expected behavior could arise from quantum tunneling between 
the qubit states. Such a tunneling process, although made small 
compared to the qubit energy splitting, violates the QND conditions.

One of the possible implementations of the system under consideration 
is the four-junction persistent current qubit \cite{Bertet05,Yoshihara06,Lupascu07} (flux qubit) 
depicted in Fig~\ref{Fig1}a). It consists of a 
superconducting loop with four Josephson junctions and its low 
temperature dynamics is confined to the two lowest-energy states. 
For an external magnetic flux close to a half-integer multiple of $\Phi_0=h/2e$, 
the superconducting flux quantum, the two lowest-energy eigenstates are 
combinations of clockwise and counter clockwise circulating current states. 
These two states represent the qubit. The measurement apparatus consists
of a superconducting quantum interference device (SQUID), composed by
two Josephson junctions, inductively coupled to the qubit loop. The SQUID 
behaves as a non-linear inductance and together with a shunt capacitance 
forms a non-linear $LC$-oscillator, which is externally driven. The two qubit 
states produce opposite magnetic field that translate into a qubit
dependent effective Josephson inductance of the SQUID. The response of
the driven SQUID is therefore qubit-dependent. 

In order to treat the problem in a fully quantum mechanical way, we linearize  
the SQUID equation of motion, such that the effective coupling between the 
driven $LC$-oscillator and the the qubit turns out to be quadratic. The qubit 
Hamiltonian is ${\cal H}_S=\epsilon\boldsymbol{\sigma}_Z/2+
\Delta\boldsymbol{\sigma}_X/2$. 
In the experiment \cite{Lupascu07}, the tunneling amplitude $\Delta$
between the two qubit current states is made small compared to the
qubit gap $E=\sqrt{\epsilon^2+\Delta^2}$, therefore also $\Delta\ll\epsilon$, such
that it can be considered as a small perturbation. The absence of the
tunneling term would yield a perfect QND Hamiltonian (see below). From the 
experimental parameters $\Delta=5~{\rm GHz}$ and $E=14.2~{\rm GHz}$
\cite{Lupascu07,Picot-arXiv08}, it follows that $\Delta/\epsilon\approx 0.38$, 
yielding a reduction of the visibility in Fig.~\ref{Fig1} d) on the order of 10\%. 

The QND character of the qubit measurement is studied by repeating 
the measurement. A perfect QND setup guarantees identical outcomes 
for the two repeated measurement with certainty. In order to fully 
characterize the properties of the measurement, we can initialize the 
qubit in the state $|0\rangle$, then rotate the qubit by applying a pulse 
of duration $\tau_1$ before the first measurement and a second pulse of duration 
$\tau_2$ between the first and the second measurement. The conditional 
probability to detect the qubit in the states $s$ and $s'$ is expected to be 
independent of the first pulse, and to show sinusoidal oscillation with 
amplitude 1 in $\tau_2$. Deviations from this expectation witness 
a deviation from a perfect QND measurement. The sequence of qubit 
pulses and oscillator driving is depicted in Fig.~\ref{Fig1}b). The conditional 
probability $P(0|0)$ to detect the qubit in the state "0" twice in sequence is 
plotted  versus $\tau_1$ and $\tau_2$ in Fig.~\ref{Fig1}c) for 
$\Delta=0$, and in Fig.~\ref{Fig1}d) for $\Delta/\epsilon=0.1$. 
We anticipate here that a dependence on $\tau_1$ is visible when the 
qubit undergoes a flip in the first rotation. Such a dependence is due to the 
imperfections of the mapping 
between the qubit state and the oscillator state, and is present also in the case 
$\Delta=0$. The effect of the non-QND term $\Delta\boldsymbol{\sigma}_X$ 
results in an overall reduction of $P(0|0)$. 

In this paper we study the effect of the tunneling term  
on the quality of a QND measurement.  
Many attempts to understand the possible origin of the deviations
from perfect QND behavior appearing in the experiments have been 
concerned with the interaction with the environment\cite{RalphPRA06,GambettaPRA06,GambettaPRA07,
SerbanPRB07,SerbanPRL07,SerbanPRB08,Picot-arXiv08}.  
The form of the Josephson non-linearity 
dictates the form of the coupling between the qubit and the 
oscillator, with the qubit coupled to the photon number 
operator of the driven harmonic oscillator, 
$\boldsymbol{\sigma}_Za^{\dag}a$, rather than to 
one quadrature, $\boldsymbol{\sigma}_X(a+a^{\dag})$, and 
the effect of the tunneling term $\boldsymbol{\sigma}_X$ 
present in the qubit Hamiltonian is considered as a small perturbation. 

The work we present is not strictly confined to the analysis of
superconducting flux qubit measurement. Rather, it is applicable 
to a generic system of coupled qubit and harmonic
oscillator that can find an application in many contexts. 
Moreover,
the analysis we present is based on the general formalism of the
positive operator valued measure (POVM), that represents the most
general tool in the study of quantum measurements.

The paper is structured as follows: in Sec.~\ref{Sec:QNDmeasurement} 
we introduce the idea of QND measurement and describe the conditions 
under which a QND measurement can be performed. In 
Sec.~\ref{Sec:QuadraticCoupling} we derive the quadratic coupling 
between the qubit and the oscillator and the Hamiltonian of the total 
coupled system. In Sec.~\ref{Sec:SingleMeasurement} we construct 
the qubit single measurement with the POVM formalism and in 
Sec.~\ref{Sec:SigmaX} we consider the effect of the non-QND term 
in the POVM that describes the single measurement.  In 
Sec.~\ref{Sec:SusequentMeasurements} we construct the two-
measurement formalism, by extending the formalism of POVM 
to the two subsequent measurement case. In 
Sec.~\ref{Sec:IdealSingleMeasurement} we consider the single measurement 
in the case $\Delta=0$ and study the condition for having a good QND 
measurement. In Sec.~\ref{Sec:OneMeasFirstOrder} we calculate the 
contribution at first order  and second order in $\Delta/\epsilon$ to the POVM 
and to the outcome probability for the qubit single measurement. In Sec.~\ref{Sec:QNDcharacter} we calculate 
the contribution at first and second order in $\Delta/\epsilon$ to the 
POVM and to the outcome probability for the qubit two subsequent 
measurements. In Sec.~\ref{Sec:Rotations} we study the QND character 
of the measurement by looking at the conditional probability for the 
outcomes of two subsequent measurements when we rotate the qubit 
before the first measurement and between the first and the second 
measurement.

\section{QND measurements}
\label{Sec:QNDmeasurement}

We consider a quantum system on which we want to measure a suitable 
observable $\hat{A}$. A measurement procedure is based on coupling the 
system under consideration to a meter. The global evolution entangles the 
meter and the system, and a measurement of an observable $\hat{B}$ of 
the meter provides information on the system. In general, a strong 
projective measurement on the meter translates into a weak non-projective 
measurement on the system. This is because the eigenstates of the 
coupled system differ in general from the product of the eigenstates 
of the measured observable on the system and those of the meter. 

Three criteria  that a measurement should satisfy in order to
be QND have been formulated \cite{RalphPRA06,Travaglione-arXiv02}: i) correct
correlation between the input state and the measurement result; ii)
the action of measuring should not alter the observable being 
measured; iii) repeated measurement should give the same
result. These three criteria can be cast in a more precise way: 
{\it the measured observable $\hat{A}$ must be an integral of motion 
for the coupled meter and system\cite{Braginsky-Khalili}}. Formally this 
means that the observable $\hat{A}$ that we want to measure 
must commute with the Hamiltonian ${\cal H}$, that describes the 
interacting system and meter,
\begin{equation}\label{Eq:QNDcondition}
[{\cal H},\hat{A}]=0.
\end{equation}  
Such a requirement represents a sufficient condition in order that an eigenstate 
of the observable $\hat{A}$, determined by the measurement, does not change
under the global evolution of the coupled system and meter. As a consequence, 
a subsequent measurement of the same observable $\hat{A}$ provides the same 
outcome as the previous one with certainty.

Finally, in order to obtain information on the system observable $\hat{A}$ 
by the measurement of the meter observable $\hat{B}$, it is necessary 
that the interaction Hamiltonian does {\it not} commute with $\hat{B}$, 
\begin{equation}
[{\cal H}_{\rm int},\hat{B}]\ne 0,
\end{equation}
where ${\cal H}_{\rm int}$ describes the interaction between the meter 
and the system, 
\begin{equation}\label{Eq:HamTotGen}
{\cal H}={\cal H}_S+{\cal H}_{\rm meter}+{\cal H}_{\rm int}. 
\end{equation}
Altogether, these criteria provide an immediate way to 
determine whether a given measurement protocol can give rise to a QND
measurement.  At this level the observables $\hat{A}$ and $\hat{B}$ and the Hamiltonian ${\cal H}$ do not pertain to any particular system. In the next section we will identify each term for the system we want to study.

\section{Model: Quadratic coupling}
\label{Sec:QuadraticCoupling}

As far as the application of our model to the measurement of a persistent 
current qubit with a SQUID is concerned, we provide here a derivation of the 
quadratic coupling mentioned in the introduction. 

We identify the system with a flux qubit that will be described by the Hamiltonian ${\cal H}_S$. The meter is represented by a SQUID and it is inductively  coupled to a flux qubit via a mutual inductance, in such a way that the qubit affects the magnetic flux through the loop of the SQUID. The Hamiltonian that describes the SQUID and the interaction with the qubit can be written as
\begin{equation}
{\cal H}_{\rm meter}+{\cal H}_{\rm int}=\frac{\hat{Q}^2}{2C}-\frac{\Phi_0^2}{L_J}
\cos\left(2\pi\hat{\Phi}/\Phi_0\right)\cos\hat{\varphi}
\end{equation}
where $\hat{\varphi}=\hat{\varphi}_1-\hat{\varphi}_2$ is the difference 
of the phases of the two Josephson 
junctions $\hat{\varphi}_1$ and $\hat{\varphi}_2$ that interrupt the SQUID 
loop, $L_J$ the Josephson inductance of the junctions (nominally equal), 
and $\hat{Q}$ is the difference of the charges accumulated on the capacitances $C$ that shunt 
the junctions. Up to a constant factor, $\hat{\varphi}$ and $\hat{Q}$ are canonically 
conjugate variables that satisfy $[\hat{\varphi},\hat{Q}]=2ei$.   
We split the external flux into a constant term and a qubit 
dependent term, such that  $\cos(2\pi\hat{\Phi}/\Phi_0)=\cos(2\pi\Phi_{\rm ext}/
\Phi_0+2\pi MI_q\boldsymbol{\sigma}_Z/\Phi_0)\equiv\lambda_0+\lambda_1
\boldsymbol{\sigma}_Z$, with $I_q$ the current in the qubit loop and $M$ 
the mutual inductance between qubit and SQUID loop. Expanding the potential 
up to second order in $\hat\varphi$, one obtains
\begin{equation}
{\cal H}_{\rm meter}+{\cal H}_{\rm int}\approx\frac{\hat{Q}^2}{2C}+
(\lambda_0+\lambda_1\boldsymbol{\sigma}_Z)
\left(\frac{\Phi_0}{2\pi}\right)^2\frac{\hat{\varphi}^2}{2L_J}.
\end{equation}
with $\lambda_0=\cos(2\pi\Phi_{\rm ext}/
\Phi_0)\cos(2\pi MI_q/\Phi_0)$ and $\lambda_1=-\sin(2\pi\Phi_{\rm ext}/
\Phi_0)\sin(2\pi MI_q/\Phi_0)$.
We introduce the zero point fluctuation amplitude $\sigma=(L_J/\lambda_0C)^{1/4}$, the bare harmonic oscillator frequency $\omega_{\rm ho}=\sqrt{\lambda_0/L_JC}$, and the in-phase and in-quadrature components of 
the field 
\begin{eqnarray}
\frac{\Phi_0}{2\pi}\hat{\varphi}\equiv\hat{X}&=&
\sigma\sqrt{\frac{\hbar}{2}}(a+a^{\dag}),\\
\hat{Q}\equiv\hat{P}&=&
-\frac{i}{\sigma}\sqrt{\frac{\hbar}{2}}(a-a^{\dag}),
\end{eqnarray}
with $a$ and $a^{\dag}$ harmonic oscillator annihilation and creation 
operators satisfying $[a,a^{\dag}]=1$. Apart from a renormalization of the qubit splitting, the Hamiltonian of the 
coupled qubit and linearized SQUID turns out to be
\begin{eqnarray}\label{Eq:QuadHam}
{\cal H}_{\rm meter}+{\cal H}_{\rm int}&=&\hbar\omega_{\rm ho}(1+\tilde g\boldsymbol{\sigma}_Z)a^{\dag}a\nonumber\\
&+&\hbar\omega_{\rm ho}\tilde g\boldsymbol{\sigma}_Z(a^2+a^{\dag}{}^2),
\end{eqnarray} 
with $\tilde g=\lambda_1/2\lambda_0=\tan(2\pi\Phi_{\rm ext}/\Phi_0)
\tan(2\pi MI_q/\Phi_0)/2$. The frequency of the harmonic 
oscillator describing the linearized SQUID is then effectively split by the qubit. 

The Hamiltonian can now be written in the form of Eq.~(\ref{Eq:HamTotGen}) with an additional driving term (from here on we set $\hbar=1$),
\begin{equation} \label{Eq:Hamiltonian}
{\cal H}(t)={\cal H}_S+{\cal H}_{\rm meter}+{\cal H}_{\rm int} +
{\cal H}_{\rm drive}(t).
\end{equation}
The qubit Hamiltonian written by means of the Pauli matrices
$\boldsymbol{\sigma}_i$ (we denote 2x2 matrices in qubit space
with bold symbols) in the basis of the current states 
$\{|0\rangle,|1\rangle\}$ is
\begin{equation}\label{Eq:HQ}
{\cal H}_S=\frac{\epsilon}{2}\boldsymbol{\sigma}_Z+
\frac{\Delta}{2}\boldsymbol{\sigma}_X,
\end{equation}
where $\epsilon=2I_q(\Phi_{\rm ext}-\Phi_0/2)$ represents an energy 
difference between the qubit states and $\Delta$ the tunneling term 
between these states. The Hamiltonian of the oscillator (or SQUID) is
\begin{equation}\label{Eq:HS}
{\cal H}_{\rm meter}=\omega_{\rm ho} a^{\dag}a.  
\end{equation}
The Hamiltonian that describes the coupling between the qubit 
and the harmonic oscillator in the rotating wave approximation (RWA), 
where we neglected the terms like $a^2$ and $a^{\dag}{}^2$, is given by 
\begin{equation}\label{Eq:HQS}
{\cal H}_{\rm int}=g \boldsymbol{\sigma}_Z a^{\dag}a,
\end{equation}
with $g=\omega_{\rm ho}\tilde{g}$ [\onlinecite{Note:int-strength}], and the 
external driving of the harmonic oscillator is described by 
\begin{equation}\label{Eq:HD}
{\cal H}_{\rm drive}(t) = f(t)(a+a^{\dag}).
\end{equation}
and throughout this work, we choose a harmonic driving force 
 $f(t)=2f\cos(\omega_dt)$. Neglecting
the fast rotating terms $ae^{-i\omega_dt}$ and
$a^{\dag}e^{i\omega_dt}$, after moving in the frame rotating with
frequency $\omega_d$, the Hamiltonian becomes time independent,
\begin{equation}\label{Eq:RWA}
{\cal H}={\cal H}_S + \Delta\boldsymbol{\omega}_Za^{\dag}a+f(a+a^{\dag}),
\end{equation}
with $\Delta\boldsymbol{\omega}_Z=\boldsymbol{\omega}_Z-\omega_d$, 
 and the qubit-dependent frequency given by
$\boldsymbol{\omega}_Z=\omega_{\rm ho}(1+\tilde{g}\boldsymbol{\sigma}_Z)$.

The qubit observable that we want to measure is 
$\hat{A}\equiv\boldsymbol{\sigma}_Z$ and, due to the presence 
of the term $\Delta\boldsymbol{\sigma}_X/2$, it does not represent 
an integral of the motion for the qubit, $[{\cal H}_S, 
\boldsymbol{\sigma}_Z]\neq 0$. Therefore the measurement 
is not supposed to be QND, Eq.~(\ref{Eq:QNDcondition}) not being satisfied. 
However, for $\Delta\ll\epsilon$ the variation in time of $\boldsymbol{\sigma}_Z$ 
becomes slow on the time scale determined by $1/\epsilon$ and one expects 
small deviations from an ideal QND case. The presence of the non-QND term
$\boldsymbol{\sigma}_X$ term in ${\cal H}_S$ inhibits an exact solution and a 
perturbative approach will be carried out in the small parameter 
$\Delta/\epsilon \ll 1$.

\section{single measurement}
\label{Sec:SingleMeasurement}

The weak measurement of the qubit is constructed as follows. 
We choose the initial density matrix $(t=0)$ of the total coupled system 
to be the product state $\boldsymbol{\rho}(0)=\boldsymbol{\rho}_0
\otimes|\hat{0}\rangle\langle\hat{0}|$, 
with the qubit in the unknown initial state $\boldsymbol{\rho}_0$ and the 
oscillator in the vacuum state $|\hat{0}\rangle$, and we let the qubit and the 
oscillator become entangled during the global time evolution. 
We then assume that at time $t$ we perform a strong measurement of the 
flux quadrature $\hat{X}=\sigma(a+a^{\dag})/\sqrt{2}$, by projecting the 
oscillator on to the state $|x\rangle\langle x|$. Such
a state of the oscillator is quite unphysical, it has infinite energy and
infinite indeterminacy of the $\hat{P}=(a-a^{\dag})/\sqrt{2}i$
quadrature. More realistically, what would happen in an experiment is
that the oscillator is projected on to a small set of quadrature states
centered around $x$. This can be described as a convolution of the projector 
$|x\rangle\langle x|$ with a distribution characteristic of the measurement 
apparatus, that can be included in the definition of the qubit weak measurement. 
However, we choose to keep the model simple and to work with an idealized projection.

In the interaction picture, the projection on the state $|x\rangle\langle x|$ corresponds to the choice to measure the quadrature 
$\hat{X}(t)=\sigma(ae^{-i\omega_{\rm ho} t}+a^{\dag}e^{i\omega_{\rm ho} t})/
\sqrt{2}$ ,
\begin{eqnarray}
x(t)&=&{\rm Tr}[\hat{X}\boldsymbol{\rho}(t)]
={\rm Tr}\left[\hat{X}(t)\boldsymbol{\rho}_R(t)\right],\\
\boldsymbol{\rho}_R(t)&=&{\cal U}_R(t)
~\boldsymbol{\rho}(0)~{\cal U}_R^{\dag}(t).
\end{eqnarray}
where an expression of ${\cal U}_R(t)$ and its derivation is given by 
Eq.~(\ref{Eq:App-UR}) in Appendix~\ref{App:Delta0}. The operator 
${\cal U}_R(t)$ describes the time-evolution of $\boldsymbol{\rho}$ 
in the rotating frame. The probability to detect  the outcome $x$ can 
then be written as  
\begin{eqnarray}
{\rm Prob}(x,t)&=&{\rm Tr}\left[\langle x|\boldsymbol{\rho}_R(t)
|x\rangle\right]\nonumber\\
&=&{\rm Tr}\left[\langle x|{\cal U}_R(t)|\hat{0}\rangle
~\boldsymbol{\rho}_0~
\langle\hat{0}|{\cal U}_R^{\dag}(t)|x\rangle\right],
\end{eqnarray}
where the trace is over the qubit space, and $\{|x\rangle\}$ is a 
basis of eigenstates of $\hat{X}(t)$. We define the operators
\begin{eqnarray}
{\bf N}(x,t) &=& \langle x|{\cal U}_R(t)|\hat{0}\rangle,\\
{\bf F}(x,t) &=& {\bf N}^{\dag}(x,t){\bf N}(x,t),
\end{eqnarray}
acting on the qubit and, using the property of invariance of the trace 
under cyclic permutation, we write
\begin{equation}\label{Eq:GenProbDef}
{\rm Prob}(x,t)={\rm Tr}~{\bf F}(x,t)
\boldsymbol{\rho}(0).
\end{equation}
The state of the system after the measurement is
$\boldsymbol{\rho}(x,t)\otimes|x\rangle\langle x|$, 
with the qubit in the state
\begin{equation}
\boldsymbol{\rho}(x,t)
=\frac{{\bf N}(x,t)\boldsymbol{\rho}(0){\bf N}^{\dag}(x,t)}{{\rm Prob}(x,t)}.
\end{equation}  
The operators ${\bf F}(x,t)$  are positive, trace- and
hermiticity-preserving superoperators (i.e. they map  
density operators into density operators) acting on the 
qubit Hilbert space. Moreover, they satisfy the 
normalization condition
\begin{equation}\label{Eq:NormContPOVM}
\int_{-\infty}^{\infty}dx{\bf F}(x,t)=\openone,
\end{equation}
from which the conservation of probability follows. Therefore, they form a 
positive operator valued measure (POVM), and we will call the operators 
${\bf F}(x,t)$ a {\em continuous} POVM. We point out here that modeling 
a more realistic scenario, by including a convolution of the projector 
$|x\rangle\langle x|$ with a distribution characteristic of the measurement 
apparatus, corresponds to the construction of a more general POVM of the 
harmonic oscillator, that would not alter qualitatively the description of the 
qubit measurement in terms of POVM.

The probability distribution ${\rm Prob}(x,t)$ depends strongly on the
initial qubit state $\boldsymbol{\rho}_0$. In general ${\rm Prob}(x,t)$ 
is expected to have a two-peak shape, arising from the two possible states of the
qubit, whose relative populations determine the relative heights of the
two peaks, one peak corresponding to $|0\rangle$ and the other to $|1\rangle$.  

We now define an indirect qubit measurement that has two possible 
outcomes, corresponding to the states ``0'' and ``1''. As a protocol for a
single-shot qubit measurement, one can measure the quadrature
$\hat{X}$ and assign the state ``0'' or ``1'' to the qubit, 
according to the two possibilities of the outcome $x$ to be greater or 
smaller than a certain threshold value $x_{\rm th}$, $x>x_{\rm th}
\rightarrow|0\rangle$ or   $x<x_{\rm th}\rightarrow|1\rangle$, as 
depicted in Fig.~\ref{Fig2}. Alternatively, 
we can infer the qubit state by repeating the procedure many times and  
constructing the statistical distribution of the outcome $x$. We then assign 
the relative populations of the qubit states $|0\rangle$  and $|1\rangle$ by 
respectively integrating the outcome distribution in the regions 
$\eta(1)=(x_{\rm th},\infty)$, $\eta(-1)=(-\infty, x_{\rm th})$. 

We formally condensate the two procedures and define a two-outcome 
POVM, that describes the two possible qubit outcomes, by writing 
\begin{eqnarray}\label{Eq:DiscreteF}
{\bf F}(s,t) &=& \int_{\eta(s)}dx{\bf F}(x,t),\qquad \\
{\rm Prob}(s,t) &=& {\rm Tr}[{\bf F}(s,t)\boldsymbol{\rho}(0)],
\end{eqnarray}
with $s=\pm 1$. We will call  ${\bf F}(s,t)$ a {\em discrete} POVM, in 
contrast to the continuous POVM ${\bf F}(x,t)$ defined above. 
Here, we introduce a convention that assigns $s=+1$ to the ``0'' 
qubit state and $s=-1$ to the ``1'' qubit state.
The probabilities ${\rm Prob}(s,t)$ are 
therefore obtained by integration of ${\rm Prob}(x,t)$ on the subsets 
$\eta(s)$, ${\rm Prob}(s,t)=\int_{\eta(s)}dx{\rm Prob}(x,t)$. On
the other hand, the probability distribution ${\rm Prob}(x,t)$ is
normalized on the whole space of outcomes which leads to
$P(0,t)+P(1,t)=1$ at all times. Typically, it is not possible to have 
a perfect mapping of the qubit state. 

\section{Effects of the tunneling $\boldsymbol{\sigma}_X$ term}
\label{Sec:SigmaX}

Deviations from an ideal QND measurement can arise due to the  
presence of a non-zero $\boldsymbol{\sigma}_X$ term in the qubit 
Hamiltonian. In SC flux qubits, such a term is usually present; it 
represents the amplitude for tunneling through the
barrier that separates the two wells of minimum potential, where the lowest
energy qubit current states are located. This term cannot be switched
off easily.

We can expand the full evolution operator 
${\cal U}_R(t)$ in powers of $\Delta t$, as in 
Eq.~(\ref{Eq:EvOp2nd-ord-exp}), and 
obtain a formally exact expansion of ${\bf F}(x,t)$,
\begin{equation}
{\bf F}(x,t)=\sum_{n=0}^{\infty}{\bf F}^{(n)}(x,t).
\end{equation} 
Due to the transverse ($X\perp Z$) character of the perturbation it follows 
that the even terms in this series (corresponding to even powers of $\Delta t$) 
have zero off-diagonal entries, whereas the odd terms have zero diagonal
entries. 
Due to the normalization condition Eq.~(\ref{Eq:NormContPOVM}),
valid at all orders in $\Delta t$, it can be shown that 
\begin{equation}\label{Eq:Nth-orderNormContPOVM}
\int dx{\bf F}^{(n)}(x,t)=\delta_{n,0}\openone, 
\end{equation}
and consequently,
\begin{equation}\label{Eq:Nth-orderNormDiscrPOVM}
\sum_{s=\pm 1}{\bf F}^{(n)}(s,t)=\delta_{n,0}\openone.
\end{equation}
As a result, the probability ${\rm Prob}(s,t)$ is given as a power
expansion in the perturbation
\begin{equation}
{\rm Prob}(s,t)=\sum_{n=0}^{\infty}{\rm Prob}^{(n)}(s,t),
\end{equation}
where $\sum_{s=\pm 1}{\rm Prob}^{(n)}(s,t)=\delta_{n,0}$. 

The expansion of the evolution operator and consequently of the continuous 
and discrete POVMs is in the parameter $\Delta t$. The requirement that the 
deviations introduced by the tunneling $\boldsymbol{\sigma}_X$ term in the 
time-evolution behave as perturbative corrections sets a time scale for the 
validity of the approximation, namely $t\ll 1/\Delta$, for which we will truncate 
the expansion up to second order. The tunneling non-QND 
term is considered as a perturbation in that experimentally one has 
$\Delta/\epsilon\ll 1$.  It turns out to be convenient to choose as a time 
scale for the qubit measurement $t\sim 1/\epsilon$, for which follows 
$\Delta t\sim \Delta/\epsilon\ll 1$.

\section{Two subsequent measurements}
\label{Sec:SusequentMeasurements}

A QND measurement implies that repeated measurements give the same 
result with certainty. In order to verify such a property of the measurement, 
we construct here the formalism that will allow us to study the correlations 
between subsequent measurements.   

After the oscillator quadrature is measured in the first step at time $t$ and
the quadrature value $x$ is detected, the total system composed of the qubit 
and the oscillator is left in the state 
$\boldsymbol{\rho}(x,t)\otimes|x\rangle\langle x|$. The fact that we can split 
the total state after the measurement into a product state is a consequence of 
the assumption that the measurement of the harmonic oscillator is a projection. 
Had a more general POVM of the harmonic oscillator be involved, then such 
a conclusion  would not hold.
After the first measurement is performed, 
the total system is left alone under the effects of dissipation
affecting the oscillator. A harmonic oscillator which is initially prepared in a 
coherent state evolves, under weak coupling to a bath of harmonic oscillators in thermal 
equilibrium, to a mixture of coherent states with a Gaussian distribution 
centered around the vacuum state (zero amplitude coherent state) with variance 
$n_{\rm th}=(\exp(\hbar\omega/k_B T)-1)^{-1}$, $\omega$ being 
the frequency of the harmonic oscillator, $T$ the temperature, and $k_B$ 
the Boltzmann constant, whereas in the case $T=0$ it evolves coherently to 
the vacuum $|\hat{0}\rangle$ [\onlinecite{Scully}].

\begin{figure}[t]
 \begin{center}
  \includegraphics[width=7cm]{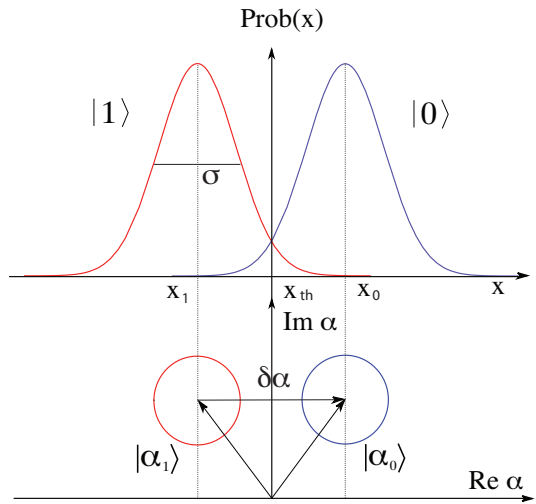}
    \caption{(Color online) Schematic description of the single measurement procedure.  In the bottom 			panel the coherent states $|\alpha_0\rangle$ and $|\alpha_1\rangle$, associated 		with the qubit states  $|0\rangle$ and $|1\rangle$, are represented for illustrative 			purposes by a contour line in the phase space at HWHM of their Wigner 				distributions, defined\cite{Scully} as $W(\alpha,\alpha^*)=(2/\pi^2)\exp(2|\alpha|^2)\int d\beta		\langle-\beta|\rho|\beta\rangle\exp(\beta\alpha^*-\beta^*\alpha)$. 
    		The corresponding Gaussian probability distributions of width $\sigma$ centered 		about the qubit-dependent "position" $x_s$ are shown in the top panel. \label{Fig2}}
 \end{center}
\end{figure}

We now assume that the 
state of the total system (qubit and oscillator) before the second measurement is
\begin{equation}
\boldsymbol{\rho}(x,t)\otimes|\hat{0}\rangle\langle\hat{0}|.
\end{equation} 
Following the previously described procedure for the qubit single-
measurement, a second measurement of the quadrature $\hat{X}$ with
outcome $y$ performed at time $t'$, having detected $x$ at time $t$,
would yield the conditional probability distribution
\begin{equation} 
{\rm Prob}(y,t'|x,t)={\rm Tr}\left[{\bf F}(y,t')
\boldsymbol{\rho}(x,t)\right].
\end{equation}
Defining the continuous POVM qubit operators for two measurements as 
\begin{equation}\label{Eq:ContDoublePOVM}
{\bf F}(y,t';x,t)={\bf N}^{\dag}(x,t)
{\bf F}(y,t'-t){\bf N}(x,t),
\end{equation}
the joint probability distributions for two subsequent measurements is
\begin{eqnarray}
{\rm Prob}(y,t';x,t)&=&{\rm Prob}(y,t'|x,t){\rm Prob}(x,t)\\
&=&{\rm Tr}\left[{\bf F}(y,t';x,t)\boldsymbol{\rho}_0\right].
\label{Eq:ProbF}
\end{eqnarray}
The operators ${\bf F}(y,t';x,t)$ satisfy the normalization condition
$\int dx\int dy{\bf F}(y,t';x,t)=\openone$, ensuring the normalization
of the probability distribution $\int dx\int dy{\rm Prob}(y,t';x,t)=1$.
By inspection of Eqs.~(\ref{Eq:NormContPOVM}) and (\ref{Eq:ContDoublePOVM}),
it follows that
\begin{equation}\label{Eq:MargProbX}
\int dy~{\bf F}(y,t';x,t)={\bf F}(x,t),
\end{equation}
and the marginal distribution for the first measurement is
\begin{equation}
{\rm Prob}_M(x,t)\equiv\int dy~{\rm Prob}(y,t';x,t)
={\rm Tr}[{\bf F}(x,t)\boldsymbol{\rho}_0],
\end{equation}
stating that the probability to detect $x$ in the first measurement
is independent on whatever could be detected in the second
measurement. On the other hand, the marginal probability distribution
for the second measurement turns out to be 
\begin{equation}
{\rm Prob}_M(y,t')\equiv\int dx{\rm Prob}(y,t';x,t)=
{\rm Tr}[{\bf F}(y,t'-t)\boldsymbol{\rho}(t)],
\end{equation}
where $\boldsymbol{\rho}(t)={\rm Tr}_S[{\cal U}_R(t)\boldsymbol{\rho}_0
\otimes|\hat{0}\rangle\langle\hat{0}|{\cal U}_R^{\dag}(t)]$ is the qubit 
reduced density matrix at time $t$. We define the discrete POVM for 
the correlated outcome measurements as
\begin{equation}
{\bf F}(s',t';s,t)=\int_{\eta(s)}dx\int_{\eta(s')}dy~
{\bf F}(y,t';x,t).
\end{equation} 

Analogously to Eq.~(\ref{Eq:MargProbX}) it follows that ${\bf F}(s,t)=\sum_{s'}{\bf
  F}(s',t';s,t)$, and the probability distribution for the outcomes of the 
  two subsequent measurement is simply given by 
\begin{equation}\label{Eq:ProbJoint}
{\rm Prob}(s',t';s,t)={\rm Tr}
[{\bf F}(s',t';s,t)\boldsymbol{\rho}_0],
\end{equation}
and it follows that
$\sum_{s'}{\rm Prob}(s',t';s,t)={\rm Prob}(s,t)={\rm Tr}[{\bf F}(s,t)
\boldsymbol{\rho}_0]$. 
The conditional probability to obtain a certain outcome $s'$ at time 
$t'$, having obtained $s$ at time $t$, is given by
\begin{equation}\label{Eq:CondProb}
{\rm Prob}(s',t'|s,t)=\frac{{\rm Tr}
[{\bf F}(s',t';s,t)\boldsymbol{\rho}_0]}{{\rm Tr}
[{\bf F}(s,t)\boldsymbol{\rho}_0]}.
\end{equation}

The discrete POVM for the double measurement can be in 
general written as
\begin{equation}\label{Eq:DefCorr}
{\bf F}(s',t';s,t)\equiv\frac{1}{2}\left[{\bf F}(s',t'){\bf F}(s,t)+h.c.\right]
+{\bf C}(s',t';s,t),
\end{equation}
where we have symmetrized the product of the two 
single-measurement discrete POVM operators ${\bf F}(s',t')$ 
and ${\bf F}(s,t)$ in order to preserve the hermiticity of each of 
the two terms of Eq.~(\ref{Eq:DefCorr}). 

Proceeding as for the case of a single qubit measurement, we expand 
${\bf F}(y,t';x,t)$ in powers of $\Delta/\epsilon$. Equating all the equal 
powers of $\Delta/\epsilon$ in the expansion it follows that 
\begin{equation}
{\bf F}^{(n)}(s,t)=\sum_{s'}{\bf F}^{(n)}(s',t';s,t),
\end{equation}
with $\sum_{ss'}{\bf F}^{(n)}(s't';s,t)=\delta_{n,0}\openone$.

\section{Ideal single measurement}
\label{Sec:IdealSingleMeasurement}

The dynamics governed by ${\cal U}_R^{(0)}(t)$ produces a coherent 
state of the oscillator, whose amplitude depends on the qubit state, see 
Fig.~\ref{Fig2}. 
In this case the continuous POVM operators have the simple form 
${\bf F}^{(0)}(x,t)=\langle\boldsymbol{\alpha}_Z(t)|x\rangle\langle x|
\boldsymbol{\alpha}_Z(t)\rangle$, defined through Eq.~(\ref{Eq:alphaZ}) 
in the Appendix \ref{App:Delta0}.  In the 
$\boldsymbol{\sigma}_Z$-diagonal basis $\{|i\rangle\}$, with $i=0,1$, 
it is given by 
\begin{equation}
{\bf F}^{(0)}(x,t)_{ij}=\delta_{ij}{\cal G}
\left(x-x_i(t)\right),
\end{equation} 
where $x_i(t)=\sqrt{2}\sigma{\rm Re}[\alpha_i(t)]$ and
${\cal G}(x)=\exp(-x^2/\sigma^2)/\sigma\sqrt{\pi}$ is a 
Gaussian of width $\sigma$ schematically depicted in 
Fig.~\ref{Fig2}. Introducing a rate 
$\kappa$ that describes the Markovian damping of the harmonic oscillator 
by a zero-temperature
bath of harmonic oscillators, the coherent state qubit-dependent
amplitude $\alpha_i(t)$ is found to be \cite{Note:coher-state-damp}
\begin{equation}
\alpha_i(t)=A_ie^{i\phi_i}\left[1-e^{-i\Delta\omega_i t-
\kappa t/2}\right],
\end{equation}
with $\Delta\omega_i=\omega_i-\omega_d$ and the qubit-dependent 
amplitudes and phases given by  
\begin{eqnarray}
A_i&=&\frac{f}{\sqrt{(\Delta\omega_i)^2+\kappa^2/4}}\label{Eq:QubDepAmp}\\
\phi_i&=&\arctan\left(\frac{\Delta\omega_i}{\kappa/2}\right)
-\frac{\pi}{2}\label{Eq:QubDepPh}.
\end{eqnarray}

The probability distribution for the $\hat{X}$ quadrature outcomes 
is then given by the sum of the two qubit-dependent Gaussians, 
weighted by the initial state state occupancy, and the discrete 
POVM for the qubit measurement as given by 
Eq.~(\ref{Eq:DiscreteF}) becomes
\begin{equation}\label{Eq:F0discPOVM}
{\bf F}^{(0)}(s,t)=\frac{1}{2}
\left[1+s~{\rm erf}\left(\frac{\delta x(t)}{\sigma}\right)
\boldsymbol{\sigma}_Z
\right],
\end{equation}
where $s=\pm 1$ labels the two possible measurement outcomes, and
$\delta x(t)=
\sigma{\rm Re}~\delta\alpha(t)/\sqrt 2$,
where $\delta\alpha(t)=\alpha_0(t)-\alpha_1(t)$.
The indirect qubit measurement gives the outcome probability 
\begin{equation}\label{Eq:POVMdelta0}
{\rm Prob}(s,t)=\frac{1}{2}
\left[1+s~{\rm erf}\left(\frac{\delta x(t)}{\sigma}\right)
\langle\boldsymbol{\sigma}_Z\rangle_0\right],
\end{equation}
with $\langle\boldsymbol{\sigma}_Z\rangle_0={\rm Tr}[\boldsymbol{\sigma}_Z\boldsymbol{\rho}_0]$. Supposing that the qubit is prepared in the $|0\rangle$ 
state, one expects to find ${\rm Prob}(0)=1$ and ${\rm Prob}(1)=0$. From 
Eq.~(\ref{Eq:POVMdelta0}), we see that even for $\Delta=0$ this is not 
always the case.

\subsection{Short time}

\begin{figure}[t]
 \begin{center}
  \includegraphics[width=8cm]{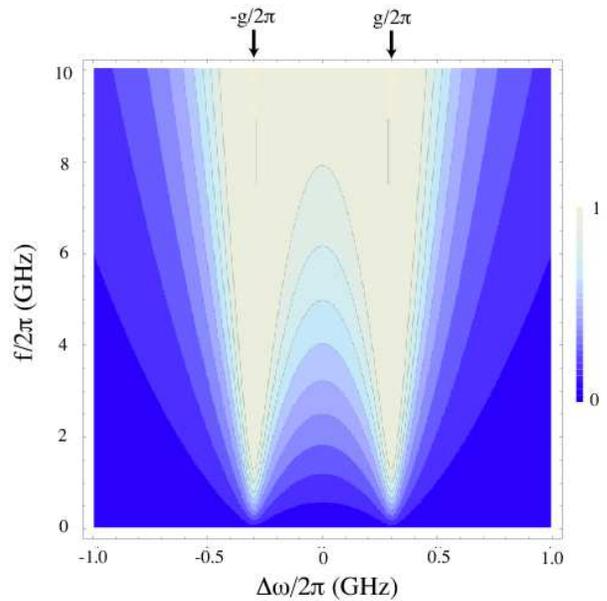}
    \caption{(Color online) ${\rm Prob}(0,t=0.1~{\rm ns})$ for the initial state
      $|0\rangle\langle 0|$, as given by Eq.~(\ref{Eq:ShTmProb}), 
      plotted as a function of the detuning $\Delta\omega/2\pi$. The values of the parameters used are listed in Tab. \ref{tab}. \label{Fig3}}
 \end{center}
\end{figure}

We choose a time $t\approx 1/\epsilon$ and a driving frequency close to the 
bare harmonic oscillator frequency. We can then expand the qubit dependent 
signal and obtain the short time behavior of the signal difference, 
\begin{equation}\label{Eq:delta-alpha-shtm}
\delta\alpha(t)\approx\sqrt{2}~t~A,
\end{equation}
with $A=f(e^{2i\phi_0}-e^{2i\phi_1})/\sqrt{2}$.
The first non-zero contribution is linear in $t$, because the signal 
is due to the time-dependent driving\cite{ClerkPRB03}. We measure a rotated quadrature
$\hat{X}_{\varphi}=\sigma(ae^{-i\varphi}+a^{\dag}e^{i\varphi})/\sqrt{2}$,
and choose the phase of the local oscillator such that 
$\varphi={\rm arg}~A$. With this choice we have 
$\delta x(t)=\sigma |A|~t$,
and the probabilities for the two measurement outcomes 
\begin{equation}\label{Eq:ShTmProb}
{\rm Prob}(s,t)=\frac{1}{2}\left[1+
s\langle\boldsymbol{\sigma}_Z\rangle_0
{\rm erf}\left(|A|~t\right)\right].
\end{equation}
In Fig.~\ref{Fig3} we plot the probability of measuring the
``0'' state ${\rm Prob}(0,t=0.1~{\rm ns})$ as a function of the detuning 
$\Delta\omega=\omega_{\rm ho}-\omega_d$ and the driving amplitude 
$f$, given that the initial state is ``0'', $\boldsymbol{\rho}_0=|0\rangle\langle 0|$.
It is possible to identify a region of values of $f$ and $\Delta\omega$ 
where  ${\rm erf}(|A| t)\approx1$ [\onlinecite{Note:erfB}].  It then follows that  
\begin{equation}
{\rm Prob}(s)\approx\frac{1}{2}\left[1+
s\langle\boldsymbol{\sigma}_Z\rangle_0\right].
\end{equation}
This case corresponds to a strong projective measurement, for which the 
outcome probabilities are either 0 ore 1, thus realizing a good qubit 
single measurement.

For driving at resonance with the bare harmonic oscillator frequency
$\omega_{\rm ho}$, the state of the qubit is encoded in the phase of the
signal, with $\phi_1=-\phi_0$, and the amplitude of the signal is actually 
reduced, as also shown in Fig.~\ref{Fig3} for $\Delta\omega=0$. When 
matching one of the two  frequencies $\omega_i$ the qubit state is encoded 
in the amplitude of the signal, as also clearly shown in Fig.~\ref{Fig3} for 
$\Delta\omega=\pm g$.  Driving away from resonance can give rise to 
significant deviation from 0 and 1 to the outcome probability, therefore 
resulting in an imprecise mapping between qubit state and measurement 
outcomes and a weak qubit measurement.

\section{Corrections due to tunneling}
\label{Sec:OneMeasFirstOrder}

\begin{figure}[t]
 \begin{center}
  \includegraphics[width=8.5cm]{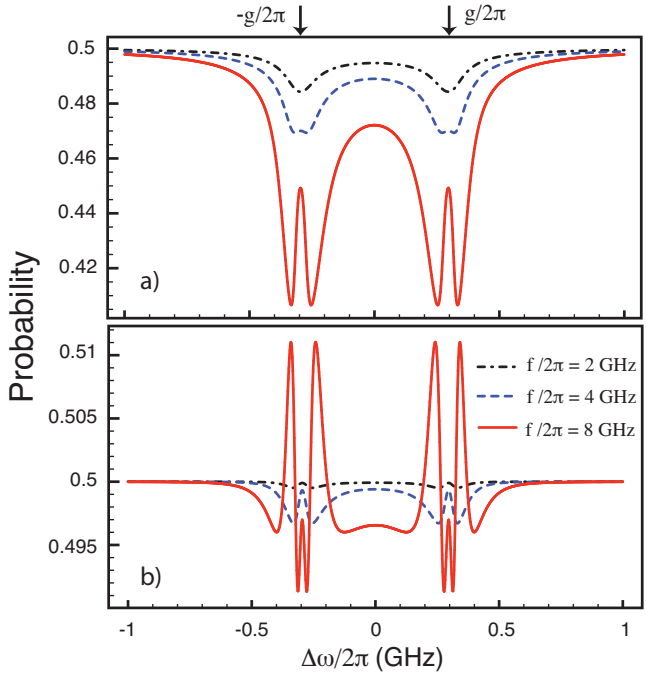}
    \caption{(Color online) Probability to detect the outcome $s=1$, corrected by a) the real 			 part of $F^{(1)}$, for the initial state $|+\rangle_X\langle +|$, and b) the 				imaginary  part of $F^{(1)}$, for the initial state $|+\rangle_Y\langle +|$, plotted 			versus the detuning $\Delta\omega/2\pi$ for several values of the amplitude $f$. 		The values of the parameters used are listed in Tab. \ref{tab}.  \label{Fig4}}
 \end{center}
\end{figure}

 \begin{figure}[t]
 \begin{center}
  \includegraphics[width=8.5cm]{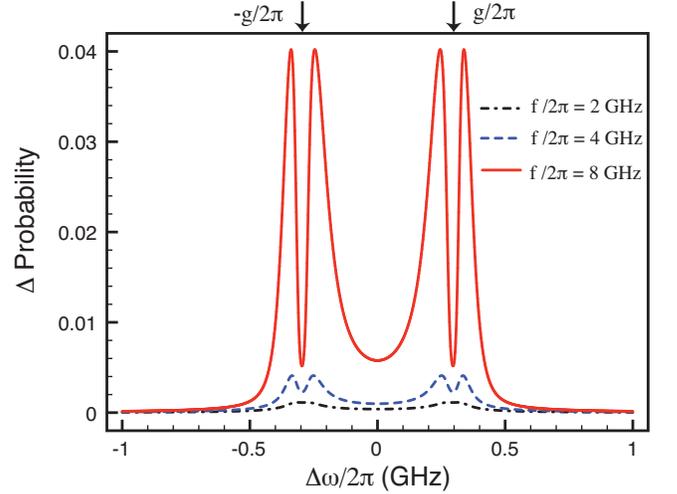}
    \caption{(Color online) Plot of the second order correction ${\rm Prob}^{(2)}(s=1)$
		 to detect "1" for the initial state 
    		$|0\rangle\langle 0|$, for $\Delta t=\Delta/\epsilon$ ,  
		as a function of the detuning 
		$\Delta\omega/2\pi$, for several values of the driving amplitude $f$. 			The values of the parameters for the evaluation  used are listed 
		in Tab. \ref{tab}.  \label{Fig5}}
 \end{center}
\end{figure}

In order to compute the correction at first order in the tunneling term 
proportional to $\Delta$ we expand the evolution operator ${\cal U}_R(t)$ 
up to first order in $\Delta t$. 
By making use of the expression Eq.~(\ref{Eq:perturbation}) for the
perturbation in the interaction picture, the off-diagonal element of
the first order correction to ${\bf F}(x,t)$ is given by 
\begin{eqnarray}
F^{(1)}(x,t)_{01}&=&-i\frac{\Delta}{2}\int_0^tdt'
\left[{\cal G}\left(x-x_0(t)+\delta z(t')\right)\right.\nonumber\\
&-&\left.{\cal G}\left(x-x_1(t)-\delta z(t')^*\right)\right]
e^{i\epsilon t'}\Gamma(t'),~~~~~~~
\end{eqnarray} 
with the complex displacement $\delta z(t)=\sigma\delta\alpha(t)/\sqrt{2}$ and  
the overlap $\Gamma(t)=\langle\alpha_0(t)|\alpha_1(t)\rangle$, where
\begin{equation}
\Gamma(t)=\exp\left(-\frac{1}{2}|\delta\alpha(t)|^2-i{\rm Im}[
\alpha_0^*(t)\alpha_1(t)]\right).
\end{equation}
Here the state ``0'' is labeled by its 
$\boldsymbol{\sigma}_Z$-eigenvalue $s=1$, whereas the state ``1'' by its
$\boldsymbol{\sigma}_Z$-eigenvalue  $s=-1$. 
Analogously to the unperturbed case, the first order contribution 
to the discrete POVM is obtained by integrating the continuous 
POVM in $x$ over the subsets $\eta(s)$. Defining the function
\begin{equation}
F^{(1)}(t)=i\frac{\Delta}{2}\int_0^tdt'
e^{i\epsilon t'}\Gamma(t')
{\rm erf}\left(\frac{\delta x(t)-\delta z(t')}{\sigma}\right),
\end{equation} 
we can write the first order contribution to the discrete POVM as 
\begin{equation}
{\bf F}^{(1)}(s,t)=s\left({\rm Re}~F^{(1)}(t)~\boldsymbol{\sigma}_X
-{\rm Im}~F^{(1)}(t)~\boldsymbol{\sigma}_Y\right),
\end{equation}
and the resulting first order correction to the probability follows directly from Eq.~(\ref{Eq:GenProbDef}). This correction is valid only for short time, $t\ll 1/\Delta$. 
For times comparable with $1/\Delta$ a perturbative expansion of 
the time evolution operator is not valid. Choosing $t\approx 1/\epsilon$, 
we can effectively approximate the phase associated with two different 
coherent states as
${\rm Im}[\alpha_0(t)\alpha_1(t)^*]\approx \psi t^2$, with 
$\psi=f^2\sin(2\phi_0-2\phi_1)$, 
the expression for $F^{(1)}(t)$ further simplifies,
\begin{equation}\label{Eq:F1}
F^{(1)}(t)=i\frac{\Delta}{2}\int_0^tdt'
e^{i\epsilon t'-\frac{1}{2}|A|^2t'^2-i\psi t'^2}
{\rm erf}\left(|A|(t-t')\right).
\end{equation} 
We study the behavior of $F^{(1)}(t)$ as a 
correction to a qubit projective measurement, that is in the range of driving 
amplitudes and frequencies that ensure ${\rm erf}(|A|~t)\approx 1$.

The real and imaginary part of $F^{(1)}(t)$ represent the first order 
correction to the outcome probability of the measurement for two 
particular initial states, respectively $|+\rangle_X\langle +|$ and 
$|+\rangle_Y\langle +|$, with 
$|\pm\rangle_X=(|0\rangle\pm|1\rangle)/\sqrt{2}$ and 
$|\pm\rangle_Y=(|0\rangle\pm i|1\rangle)/\sqrt{2}$. In the first 
case we have 
\begin{equation}
{\rm Prob}(s,t)=\frac{1}{2}+s~{\rm Re}F^{(1)}(t),
\end{equation}
and analogously for the second case, with the imaginary part 
instead of the real one. We see that the probability to obtain "0" 
is increased by ${\rm Re}F^{(1)}(t)$ and the 
probability to obtain "1" is decreased by the same amount. 
Since the contribution to first order in $\Delta t$ only affects
the off-diagonal elements of $\boldsymbol{\rho}_0$, there is no effect, 
at first order for the qubit basis states $|0\rangle$ and $|1\rangle$. 

In Fig.~\ref{Fig4} a) we plot the probability to detect the outcome state "0", 
corresponding to the outcome $s=1$, corrected up to first order in the 
perturbation for $\Delta t=\Delta/\epsilon=0.1$, for the initial state 
$\boldsymbol{\rho}_0=|+\rangle_X\langle+|$, that involves 
${\rm Re}~F^{(1)}(t)$. We see that the effect of the tunneling is largest when 
driving at resonance with the two qubit-shifted frequencies, 
$\Delta\omega\approx\pm g$. For weak driving amplitude $f$, the phase $\psi$ in Eq.~(\ref{Eq:F1}) is small and the response is of order of $\sim1\%$, close to the qubit-split frequency. By increasing the strength of the driving we see that the structure acquires two local minima in proximity of the resonance  $\Delta\omega\approx\pm g$ and a maximum exactly at resonance $\Delta\omega=\pm g$. The strong oscillatory behavior of the probability  is due to a rapid change of sign of the phase $\psi$ in proximity of the qubit-split frequencies, that is enhanced when the driving strength $f$ increases. 
In Fig.~\ref{Fig4} b) we plot the probability to detect the outcome state "0", 
corresponding to the outcome $s=1$ for the initial state 
$\boldsymbol{\rho}_0=|+\rangle_Y\langle+|$, that 
involves ${\rm Im}~F^{(1)}(t)$. In comparison to Fig.~\ref{Fig4} a) we find twice as many oscillations in the structure, typical for the imaginary part of a response function, 
when compared to the real part, and an overall scale factor of order 0.1. Besides, the sign of the response is not unique. The scale factor and the sign are understood from the Hamiltonian of the qubit, ${\cal H}=(\epsilon\boldsymbol{\sigma}_Z+\Delta\boldsymbol{\sigma}_X)/2$, with the condition $\Delta\ll\epsilon$. Under free evolution for a time $t\sim 1/\epsilon$, the initial state $|+\rangle_X\langle+|$  acquires a larger component in the $Z$-direction than the initial state $|+\rangle_Y\langle+|$. Deviation from this naive picture due to the coupling with the measurement apparatus translates in fluctuations that may determine a change of sign in the response for the case $\boldsymbol{\rho}_0=|+\rangle_Y\langle+|$.
Away from the resonances we see no significant contribution to the 
outcome probability.

\begin{table}[b]
\begin{tabular}{c|c|c}\hline\hline
Quantity & ~~~Symbol~~~ & ~~~Value for plots~~~\\
\hline\hline
Qubit detuning & $\epsilon$ & $2\pi\times 10~$GHz\\
Damping rate & $\kappa$ & $2\pi\times 0.1~$GHz\\
Coupling strength & $g$ & $2\pi\times 0.3~$GHz\\
Qubit tunneling & $\Delta/\epsilon$ & $0.1$\\
\hline\hline
\end{tabular}
\caption{Values of the parameters used in the plots. \label{tab}} 
\end{table}

First order effects in the tunneling cannot be 
responsible for qubit flip during the measurement. 
In order to estimate the deviation from a perfect QND measurement for 
the eigenstates of $\boldsymbol{\sigma}_Z$, we have to consider the 
effect of the perturbation at second order.
We define $F^{(2)}(t)$ in Eq.~(\ref{Eq:F2}) and the contribution at 
second order in $\Delta t$ to the discrete POVM is then
\begin{equation}
{\bf F}^{(2)}(s,t)=-sF^{(2)}(t)~\boldsymbol{\sigma}_Z.
\end{equation}
The dependence on $s$ factorizes, as expected from the symmetry 
between the states $|0\rangle$ and $|1\rangle$, in the picture we 
consider with no relaxation mechanism. The correction at second 
order in $\Delta/\epsilon$ to the outcomes probability is given by
\begin{equation}
{\rm Prob}^{(2)}(s,t)=-sF^{(2)}(t)\langle\boldsymbol{\sigma}_Z\rangle_0.
\end{equation}
In Fig.~\ref{Fig5} we plot the second order correction to the probability to 
obtain "1" having prepared the qubit in the initial state $\boldsymbol{\rho}_0=|0\rangle\langle 0|$, corresponding to $F^{(2)}(t)$, for $\Delta t=\Delta/\epsilon=0.1$. We choose to plot only the deviation from the unperturbed probability  because we want to highlight the contribution to spin-flip purely due to  tunneling in the qubit Hamiltonian. In fact most of the contribution to spin-flip arises from the unperturbed probability, as it is clear from Fig.~\ref{Fig3}. Around the two qubit-shifted frequencies, the probability has a two-peak structure whose characteristics come entirely from the behavior of the phase $\psi$ around the resonances $\Delta\omega\approx \pm g$. We note that the tunneling term can be responsible for a probability correction up to $\sim 4\%$ around the qubit-shifted frequency.

From the analysis of the qubit single measurement in Fig.~\ref{Fig3} we conclude 
that a weak POVM qubit measurement, that yields a large error in the determination of the qubit state, can arise when weakly driving the harmonic oscillator. Therefore, only a strong qubit projective measurement, obtained for strong driving of the oscillator, can produce a confident mapping of the qubit state at the level of a single measurement. In this case, a deviation on the order of a few percent in the state assignment can be ascribable to the tunneling term.

\section{QND character of the qubit measurement}
\label{Sec:QNDcharacter} 

As explained in Sec.~\ref{Sec:QNDmeasurement}, repeated 
measurements should give the same result if the measurement is QND. 
Such a requirement means that if a measurement projects the system 
onto an eigenstate of the measured observable, then a subsequent 
measurement should give the same result with certainty. The presence 
of a term that does not satisfy the QND condition may affect the character 
of the measurement essentially in two ways: i) by introducing deviations 
from the projection character of the single measurement, and ii) by 
generating non-zero commutators in the two-measurement POVM.
These may affect the two-outcome probabilities.

\subsection{$\Delta=0$ case}

The case $\Delta=0$ satisfies the requirement for a QND 
measurement of the qubit observable $\boldsymbol{\sigma}_Z$. 
The discrete POVM factorizes in this particular case, by virtue of 
the fact that $[{\bf N}^{(0)}(y,t'-t),{\bf N}^{(0)}(x,t)]=0$, 
\begin{equation}
{\bf F}^{(0)}(s',t';s,t)={\bf F}^{(0)}(s',t'-t){\bf F}^{(0)}(s,t).
\end{equation}
Choosing, e.g., $t'=2t$ and using Eq.~(\ref{Eq:ProbJoint}), the joint probability 
for the two measurements reads
\begin{eqnarray}
{\rm Prob}(s';s)&=&\frac{1}{4}\left[1+s's~{\rm erf}
\left(\frac{\delta x(t)}{\sigma}\right)^2\right.\nonumber\\
&+&\left.(s'+s){\rm erf}\left(\frac{\delta x(t)}{\sigma}\right)
\langle\boldsymbol{\sigma}_Z\rangle_0\right].
\end{eqnarray}
In the region of driving frequency and amplitude that ensure 
${\rm erf}(\delta x/\sigma)\approx 1$, we find 
\begin{equation}\label{Eq:Pcond00proj}
{\rm Prob}(s'|s)=\frac{1+s's+(s+s')\langle\boldsymbol{\sigma}_Z\rangle_0}
{2(1+s\langle\boldsymbol{\sigma}_Z\rangle_0)},
\end{equation}
with ${\rm Prob}(s;s)={\rm Prob}(s)$, 
and ${\rm Prob}(-s;s)=0$, and the conditional probability is ${\rm Prob}(s|s)=1$, and ${\rm Prob}(-s|s)=0$, 
regardless of $\langle\boldsymbol{\sigma}_Z\rangle_0$. However, 
it has to be noticed that in the case the condition 
${\rm erf}(\delta x/\sigma)\approx 1$ does not perfectly hold, the 
conditional probability for the two measurement to give the same outcome 
becomes
\begin{equation}\label{Eq:Delta0Pcond00}
{\rm Prob}(s|s)=\frac{1+{\rm erf}(\delta x/\sigma)^2+2s~
{\rm erf}(\delta x/\sigma)\langle\boldsymbol{\sigma}_Z\rangle_0}
{2(1+s~{\rm erf}(\delta x/\sigma)\langle\boldsymbol{\sigma}_Z\rangle_0)},
\end{equation}
and this does depend on the initial state 
$\langle\boldsymbol{\sigma}_Z\rangle_0$.

\subsection{First order contribution}

We now apply the perturbative approach in $\Delta t$ to 
estimate the effect of the non-QND term for the joint and the 
conditional probabilities. Due to the transverse nature of the 
perturbation, it is possible to show that all the odd terms have 
off-diagonal entries, whereas even ones are diagonal. 
At first order in $\Delta t$ the off-diagonal term of the 
discrete POVM is given by
\begin{equation}
F^{(1)}(s',t';s,t)=
\frac{s}{2}F^{(1)}(t)+
\frac{s'}{2}F^{(1)}(t'-t)
+s'C^{(1)}(t';t),
\end{equation}
with the quantity $C^{(1)}(t';t)$ given by Eq.~(\ref{EqApp:C1}) in 
Appendix \ref{App:SecOrder}. 
For the particular choice $t'=2t$, for which the two measurement 
procedures are exactly the same, the joint probability for the initial 
state $\boldsymbol{\rho}_0=|+\rangle_X\langle+|$ is given at first 
order in $\Delta t$ by
\begin{eqnarray}
{\rm Prob}(s',s)&=&\frac{1}{4}\left(1+s's~{\rm erf}\left(\frac{\delta x(t)}
{\sigma}\right)^2\right)\nonumber\\
&+&\frac{1}{2}(s+s'){\rm Re}~F^{(1)}(t)
+s'~{\rm Re}~C^{(1)}(2t;t)\nonumber\\
\end{eqnarray} 
We immediately observe that the probability is not symmetric with 
respect to $s$ and $s'$. Although the driving times are the same, 
something is different between the first and the second measurement, 
and the probability to obtain different outcomes $s'=-s$ is different from zero.
An analogous result holds for the initial state $\boldsymbol{\rho}_0
=|+\rangle_Y\langle+|$, with the imaginary part instead of the real one. 
Now, no matter the sign of $C^{(1)}$, the product $-s~C^{(1)}$ is 
negative in one case ($s=\pm 1$). In order to ensure that probabilities 
are non-negative one has to choose $\Delta t$ small enough such that 
the first order negative corrections due to $C^{(1)}$ remains smaller than 
the unperturbed probability. If $\Delta t$ is too large, one needs to take 
higher orders into account which should then ensure an overall non-
negative probability. The behavior of  $C^{(1)}$ as a function of the 
detuning $\Delta\omega$ and the driving amplitude $f$ is very similar 
to that of $F^{(1)}$ and we choose not to display it. The only main 
difference arises in the magnitude, for which we have $|C^{(1)}|\ll|F^{(1)}|$. 
It is the clear that the main deviations in the two measurement probabilities 
are mainly due to the errors in the first or second measurement.

\subsection{Second order contribution}

The contribution to the discrete POVM at second order in
$\Delta t$ can be divided into a term that factorizes
the contributions of the first and the second measurements, 
as well as a term that contains all the non-zero commutators 
produced in the rearrangement,
\begin{eqnarray}
{\bf F}^{(2)}(s',t';s,t)&=&{\bf F}^{(0)}(s,t){\bf F}^{(2)}(s',t'-t)
\nonumber\\
&+&{\bf F}^{(2)}(s,t){\bf F}^{(0)}(s',t'-t)\nonumber\\
&+&\frac{1}{2}\left[{\bf F}^{(1)}(s,t){\bf F}^{(1)}(s',t'-t)
+h.c.\right]\nonumber\\
&+&{\bf C}^{(2)}(s',t';s,t).
\end{eqnarray}
The full expression of the ${\bf C}^{(2)}$ at second order is rather involved. 
Choosing $t'=2t$ we then obtain
\begin{equation}\label{Eq:C2}
C^{(2)}(p',2t;p,t)_{ss}=p'ps~C^{(2)}(t)
-p'p\left|F^{(1)}(t)\right|^2,
\end{equation}
with $C^{(2)}(t)$ given by Eq.~(\ref{Eq:C2app}) in Appendix~\ref{App:SecOrder}. 
The probability to obtain identical outcomes does depend on the outcome $s$ 
itself, and this reflects the fact that the joint probability still depends on the 
initial states of the qubit. At the same time, the probability for obtaining different 
outcomes does not depend on $s$, as expected. However, direct evaluation 
of the function $C^{(2)}(p',2t;p,t)$ shows that its contribution to the probability 
is of order $0.1\%$ and can be neglected.

\section{Rabi oscillations between measurements}
\label{Sec:Rotations}

\begin{figure}[t]
 \begin{center}
  \includegraphics[width=9cm]{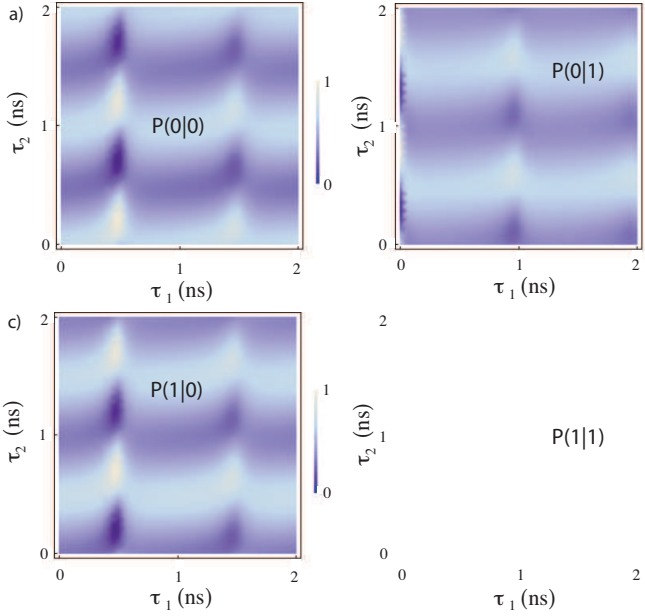}
    \caption{(Color online) Conditional probability to obtain a) $s'=s=1$, b) $s'=-s=1$, 
    		c) $s'=-s=-1$, and d) $s'=s=-1$ for the case $\Delta t=\Delta/\epsilon=0.1$ and $T_1=10~{\rm ns}$, 
		when rotating the qubit around the $y$ axis before the first measurement 
		for a time $\tau_1$ and between the first and the second measurement 			for a time $\tau_2$, starting with the qubit in the state 
		$|0\rangle\langle 0|$. Correction in $\Delta t$ 
		are up to second order. The harmonic oscillator is driven at resonance 
		with the bare harmonic frequency and a strong driving together with a strong damping of the oscillator are assumed, with $f/2\pi=20~{\rm GHz}$ and $\kappa/2\pi=1.5~{\rm GHz}$.  \label{Fig6}}
 \end{center}
\end{figure}

\begin{figure}[t]
 \begin{center}
  \includegraphics[width=9cm]{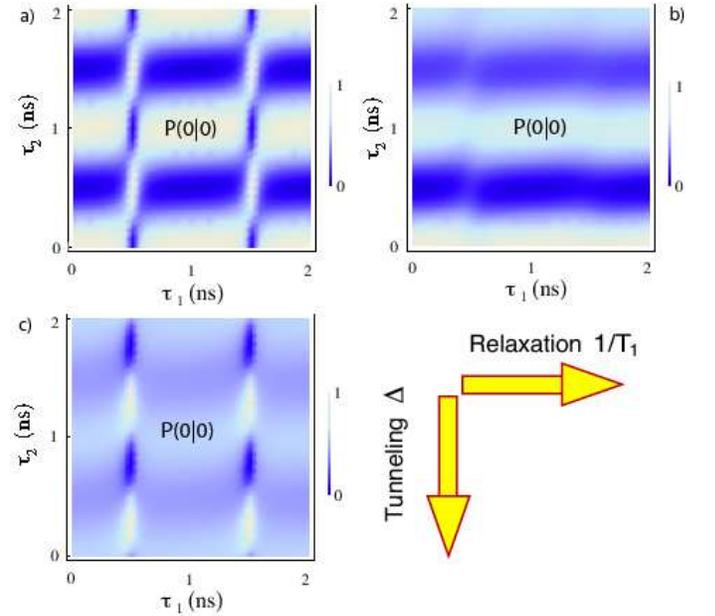}
    \caption{(Color online) Comparison of the deviations from QND behavior originating from different mechanisms. Conditional probability $P(0|0)$ versus qubit driving time $\tau_1$ and $\tau_2$ starting with the qubit in the state $|0\rangle\langle 0|$, for a) $\Delta=0$ and $T_1=\infty$, b)  $\Delta=0$ and $T_1=2~{\rm ns}$, and c) $\Delta=0.1~\epsilon$ and $T_1=\infty$. The oscillator driving amplitude is $f/2\pi=20~{\rm GHz}$ and a damping rate $\kappa/2\pi=1.5~{\rm GHz}$ is assumed. \label{Fig7}}
 \end{center}
\end{figure}

In order to gain a full insight in the QND character of the measurement, we 
analyze the behavior of the conditional probability to detect the outcomes 
$s$ and $s'$ in two subsequent measurements when we perform a rotation 
of the qubit between the two measurements. Such a procedure has been 
experimentally adopted in the work of Lupa\c{s}cu {\it et al.} [\onlinecite
{Lupascu07}].  When changing the qubit state between the two measurements, 
only partial QND behavior is expected. In addition to this, we apply an initial 
rotation to the qubit, such that a wide spectrum of initial states is tested. 
Ideally, the complete response of this procedure is supposed to be 
independent on the time $\tau_1$, during which we rotate the qubit before 
the first measurement, and to depend only on the time $\tau_2$, during which 
we rotate the qubit between the first and the second measurements, 
with probabilities ranging from zero to one as a function of $\tau_2$. 
Such a prediction, once confirmed, would guarantee a full QND character 
of the measurement.  

In Fig.~\ref{Fig1} c) we plot the conditional probability 
$P(0|0)$ for the case $\Delta=0$, when strongly driving the harmonic oscillator at resonance with the bare harmonic frequency, $\Delta\omega=0$. 
The initial qubit state is chosen to be $|0\rangle\langle 0|$. No dependence 
on $\tau_1$ appears and the outcomes $s$ and $s'$ play a symmetric role.  
This is indeed what we expect from a perfect QND measurement. 
In Fig.~\ref{Fig6} we plot the four combinations of conditional probability 
$P(s'|s)$ up to second order corrections in $\Delta t=\Delta/\epsilon=0.1$ and 
with a phenomenological qubit relaxation time $T_1=10~{\rm ns}$. 
We choose $\Delta\omega=0$, that is at resonance with the bare harmonic 
frequency. The initial qubit state is $|0\rangle\langle 0|$.  
Three features appear: i) a global reduction of the visibility of the oscillations, 
ii) a strong dependence on $\tau_1$  when the qubit is completely flipped in the 
first rotation and iii) an asymmetry under change of the outcome of the first 
measurement, with an enhanced reduction of the visibility when the first 
measurement produces a result that is opposite with respect to the initial 
qubit preparation $|0\rangle\langle 0|$. Furthermore, we find a weak 
dependence of the visibility on $\tau_1$.

We now investigate whether it is possible to identify the contribution of different 
mechanisms that generate deviations from a perfect QND measurement. In 
Fig.~\ref{Fig7} we study separately the effect of qubit relaxation and qubit 
tunneling on the conditional probability $P(0|0)$. In Fig.~\ref{Fig7} a) we set 
$\Delta=0$ and $T_1=\infty$. The main feature appearing is a sudden change 
of the conditional probability $P\rightarrow 1-P$ when the qubit is flipped in the 
first rotation. This is due to imperfection in the mapping between the qubit state 
and the state of the harmonic oscillator, already at the level of a single 
measurement. The inclusion of a phenomenological qubit relaxation time 
$T_1=2~{\rm ns}$, intentionally chosen very short, yields a strong damping of the oscillation along $\tau_2$ and washes out the response change when the 
qubit is flipped during the first rotation. This is shown in Fig.~\ref{Fig7} b). The 
manifestation of the non-QND term comes as a global reduction of the visibility 
of the oscillations, as clearly shown in Fig.~\ref{Fig7} c). 

At this level it is clearly possible to associate the observed features to different 
originating mechanisms: i) qubit tunneling yields an overall reduction of the 
visibility of the oscillations and an asymmetry under change of the outcome 
of the first measurement qubit relaxation and deviation from projective 
measurements, ii) qubit relaxation results in damping along $\tau_2$ and weak 
dependence of the oscillations on $\tau_1$, and iii) deviations from projective measurements show up mostly when the qubit is flipped during the first rotation.

The combined effect of the quantum fluctuations of the 
oscillator together with the tunneling between the qubit states is therefore 
responsible for deviation from a perfect QND behavior, although a major 
role is played, as expected, by the non-QND tunneling term.  Such a 
conclusion pertains to a model in which the qubit QND measurement is 
studied in the regime of strong projective qubit measurement and  
qubit relaxation is taken into account only phenomenologically. 
We compared the conditional probabilities plotted in Fig.~\ref{Fig6} and  
Fig.~\ref{Fig7} directly to Fig. 4 in Ref.~[\onlinecite{Lupascu07}], where 
the data are corrected by taking into account qubit relaxation, and find 
good qualitative agreement. 

Our findings can also be compared to the experiment [\onlinecite{BoulantPRB07}], in which the QND character of the measurement is 
addressed by studying a series of two subsequent measurements, but no 
qubit rotation is performed between the two measurements. 
The data in Ref.~[\onlinecite{BoulantPRB07}] are affected by strong qubit relaxation. However, from the analysis of the joint probabilities of the outcomes of the two measurements provided in Ref.~[\onlinecite{BoulantPRB07}], one can extract the conditional probabilities $P(0|0)\sim 83\%$ (when starting with the qubit initially in the ground state and comparable to Fig.~\ref{Fig6} a) at 
$\tau_1=\tau_2=0$), and 
$P(0|0)\sim 77\%$ (after a $\pi$-pulse is applied to the qubit initially in the ground state, that is comparable to Fig.~\ref{Fig6} a) at $\tau_1=0.5~{\rm ns}$ and $\tau_2=0$). In these cases one would expect a conditional probability of order 1 and a weak dependence on qubit relaxation. A deviation of order $\sim 20\%$ can be understood within the framework of our model as arising from the non-QND term and from a weak qubit measurement. Besides, from the data provided in Ref.~[\onlinecite{BoulantPRB07}], one can extract a probability of $\sim 17\%$ to obtain the excited state, when starting with the qubit in the ground state, already at the level of the single measurement. Such a behavior cannot be understood as a result of qubit relaxation and it can be ascribed to deviations from a projective qubit measurement.

\section{Conclusion}

In this paper we have analyzed the QND character of a qubit measurement 
based on coupling to a harmonic oscillator that works as a pointer to the qubit 
states. The Hamiltonian that describes the interaction between the qubit and 
the oscillator does not commute with the qubit Hamiltonian. This would in 
principle inhibit a QND measurement of the qubit. The term in the qubit 
Hamiltonian that gives rise to the non-zero commutator is small compared 
with the qubit energy gap, and in the short time qubit dynamics it can be 
viewed as a small perturbation. The perturbative analysis carried out for fast measurements leads us to the conclusion that the effect of the non-QND 
term can manifest itself as a non negligible correction. A perfect QND 
measurement guarantees perfect correlations in the outcomes of two 
subsequent measurements, therefore QND character of the measurement 
is understood in terms of deviations from the expected behavior. Corrections 
to the outcome probabilities have been calculated up to second order in the 
perturbing term.

The ground and excited states of the qubit  are affected only 
at second order by the perturbation, but a general measurement 
protocol should prescind from the state being measured. Therefore, in 
the spirit of the experiment of Lupa\c{s}cu {\it et al.} [\onlinecite{Lupascu07}], 
we have studied the conditional probability for the outcomes of two subsequent 
measurements when rotating the qubit before the first measurement 
and between the first and the second measurement. In the case where 
the QND condition is perfectly satisfied, that is when the perturbation is 
switched off, we see no dependence of the conditional probability on the 
duration of the first rotation appears and the Rabi oscillations 
between the two measurement range from zero to one. This behavior 
shows perfect QND character of the qubit measurement. On the other 
hand, the main effects of the non-QND term manifests itself as an overall 
reduction of the visibility of the oscillations and a asymmetry between 
the outcomes of the measurements. An additional dependence on the 
duration of the first qubit rotation may appear if a projective 
measurement of the qubit is not achieved already in absence of the 
perturbing non-QND term.  Experimentally the measurement is not 
projective and relaxation processes inhibit a perfect flip of the qubit 
before the first measurement. 

We point out that our analysis is valid only when the non-QND term 
$\Delta\boldsymbol{\sigma}_X$ can be viewed as a perturbation, that is for 
short time $\Delta t\ll 1$ and when the qubit dynamics is dominated by the 
term $\epsilon\boldsymbol{\sigma}_Z$, for $\Delta/\epsilon\ll 1$. 
Our analysis is not valid for the case $\epsilon=0$. In the present study we have neglected the non-linear character of the SQUID, which is not relevant to the fundamental issue described here, but plays an important role in some 
measurement procedures\cite{SiddiqiPRL04,SiddiqiPRB06,Lupascu07,
BoulantPRB07}.

A way to improve the QND efficiency would be simply to switch the 
tunneling off. In the case of superconducting flux qubit, a possibility toward 
smaller $\Delta$ could be to gate the superconducting islands between the  
junctions of the qubit loop [\onlinecite{Chirolli-PRB06}]. As an operational 
scheme one could think of working at finite $\Delta$ for logical operations 
and then at $\Delta=0$ for the measurement.

We acknowledge funding from the DFG within SPP 1285 "Spintronics"
and from the Swiss SNF via grant no. PP02-106310.

\appendix

\section{Exactly solvable case: $\Delta=0$}
\label{App:Delta0}

In order to determine the evolution governed by the Hamiltonian 
Eq.~(\ref{Eq:RWA}) we single out the term ${\cal H}_0$ diagonal in the 
$\{|s,n\rangle\}$ basis, with $|s\rangle$ the eigenstates of 
$\boldsymbol{\sigma}_Z$ and $|n\rangle$ the oscillator Fock states, 
\begin{equation}
{\cal H}={\cal H}_0+f(a+a^{\dag})+\frac{\Delta}{2}\boldsymbol{\sigma}_X,
\end{equation} 
with ${\cal H}_0=\epsilon\boldsymbol{\sigma}_Z/2+
\Delta\boldsymbol{\omega}_Za^{\dag}a$. We then work in the interaction 
picture with respect to ${\cal H}_0$. The Heisenberg equation for the density 
operator reads $\dot{\boldsymbol{\rho}}_I=-i\left[{\cal H}_I,
\boldsymbol{\rho}_I\right]$, with 
\begin{eqnarray}\label{Eq:IntPicHam}
{\cal H}_I&=&{\cal H}_I^{(0)}+V_I,\\
{\cal H}_I^{(0)}&=&
f(ae^{-i\Delta\boldsymbol{\omega}_Zt}+a^{\dag}e^{i\Delta\boldsymbol{\omega}_Zt}),\\
V_I&=&\frac{\Delta}{2}\left(e^{i\hat{\Omega}_nt}\boldsymbol{\sigma}_++
e^{-i\hat{\Omega}_nt}\boldsymbol{\sigma}_-\right),
\end{eqnarray}
where we define $\hat{\Omega}_n=\epsilon+2ga^{\dag}a$, and  
$\boldsymbol{\sigma}_{\pm}=(\boldsymbol{\sigma}_X\pm
i\boldsymbol{\sigma}_Y)/2$. We will call ${\cal U}_I(t)$ the evolution operator 
generated by ${\cal H}_I$. 

The evolution operator is given by 
${\cal U}(t)=\exp(-i\omega_dt a^{\dag}a-i{\cal H}_0 t){\cal U}_I(t)$.
For the measurement procedure so far defined we are interested in the 
evolution operator in the frame rotation at the bare harmonic oscillator 
frequency. Therefore 
\begin{equation}\label{Eq:App-UR}
{\cal U}_R(t)=\exp(-i\epsilon t\boldsymbol{\sigma}_Z/2
-i{\cal H}_{\rm int} t){\cal U}_I(t). 
\end{equation}

For the case $\Delta=0$ the model is exactly solvable and 
${\cal U}^{(0)}_I(t)$ can be computed via a generalization 
of the Baker-Hausdorff formula \cite{Gardiner},
\begin{eqnarray}\label{Eq:Ev0th}
{\cal U}_I^{(0)}(t)=D(\boldsymbol{\gamma}_Z(t)),
\end{eqnarray}
with the qubit-dependet amplitude 
$\boldsymbol{\gamma}_Z(t)=-if
\int_0^tdse^{i\Delta\boldsymbol{\omega}_Zs}$. 
The operator $D(\alpha)={\rm exp}(a^{\dag}\alpha-a\alpha^*)$ is a displacement
operator \cite{Scully}, and it generates a coherent state when applied to the vacuum 
$|\alpha\rangle\equiv D(\alpha)|0\rangle=e^{-|\alpha|^2/2}\sum_n(\alpha^n/\sqrt{n!})|n\rangle$.
In the frame rotating at the bare harmonic oscillator frequency, the state of 
the oscillator is a coherent state whose amplitude depends on the qubit state. 
A general initial state 
\begin{equation}\label{Eq:ProdState}
\boldsymbol{\rho}_{\rm tot}(0)=
\sum_{ij=0,1}\rho_{ij}|i\rangle\langle j|\otimes
|\hat{0}\rangle\langle\hat{0}|,
\end{equation}
where $|\hat{0}\rangle$ is the harmonic oscillator vacuum state, evolves to
\begin{equation}
\boldsymbol{\rho}_R(t)=\sum_{ij=0,1}\rho_{ij}|i\rangle\langle j|\otimes
|\alpha_i(t)\rangle\langle \alpha_j(t)|,
\end{equation}
where we define the qubit operators $\boldsymbol{\alpha}_Z(t)\equiv
\boldsymbol{\gamma}_Z(t)e^{-igt\boldsymbol{\sigma}_Z}$, and the object
\begin{equation}\label{Eq:alphaZ}
|\boldsymbol{\alpha}_Z(t)\rangle\equiv
D(\boldsymbol{\alpha}_Z)|\hat{0}\rangle,
\end{equation}
that gives a qubit-dependent coherent state of the harmonic oscillator,
once the expectation value on a qubit state is taken, 
$|\alpha_i(t)\rangle=\langle i|\boldsymbol{\alpha}_Z(t)|i\rangle$, 
for $i=0,1$.

\subsection{Perturbation theory in $\Delta$}
\label{App:Delta}

For non-zero $\Delta$, a formally exact solution can be written as
\begin{equation}
{\cal U}_I(t)={\cal U}_I^{(0)}(t){\cal T}\exp\left(-i~
\Delta\int_0^tdt'{\cal V}_I(t')\right),
\end{equation}
with ${\cal V}_I(t)={\cal U}_I^{(0)}{}^{\dag}(t)V_I(t){\cal U}_I^{(0)}(t)$ 
and ${\cal T}$ the time order operator. For a time scale $t\ll 1/\Delta$ 
we expand the evolution operator in powers of $\Delta t\ll 1$ ,
\begin{eqnarray}\label {Eq:EvOp2nd-ord-exp}
{\cal U}_I(t)&\approx&{\cal U}_I^{(0)}(t)
\left(\openone-i\Delta t\int_0^1ds {\cal V}_I(s~t)\right.\nonumber\\
&-&\left.(\Delta t)^2\int_0^1ds\int_0^{s}ds' {\cal V}_I(s~t){\cal V}_I
(s'~t)\right).~~
\end{eqnarray}
The interaction picture potential can be  written as 
\begin{equation}\label{Eq:perturbation}
{\cal V}_I(t)=\frac{1}{2}\left[{\cal D}(t)\boldsymbol{\sigma}^++
{\cal D}^{\dag}(t)\boldsymbol{\sigma}^-\right],
\end{equation} 
with the oscillator operators ${\cal D}(t)$ defined as
\begin{eqnarray}
{\cal D}(t)&=&D^{\dag}(\gamma_0(t))e^{i\Omega_nt}D(\gamma_1(t))\\
&=&\exp\left(i\epsilon t-i{\rm Im}[\alpha_0(t)\alpha_1(t)^*]\right)\nonumber\\
&\times&
D(-\delta\alpha(t)e^{igt})e^{2igta^{\dag}a}.
\end{eqnarray}
Here $\delta\alpha(t)=\alpha_0(t)-\alpha_1(t)$ is the difference between
the amplitudes of the coherent states associated with the two possible
qubit states.

\section{First and Second order quantities $C^{(1)}$, $F^{(2)}$ and $C^{(2)}$}
\label{App:SecOrder}

For time $t\approx 1/\epsilon$, we expand the evolution operator in 
$\Delta t$ and collect the contributions that arise at second power 
in $(\Delta/\epsilon)$. 
By making use of the expression Eq.~(\ref{Eq:perturbation}) for the
perturbation in the interaction picture we can compute the qubit 
components of the second order contribution to the continuous POVM.  
We define
\begin{eqnarray}
{\cal O}_s(t',t'')&=&\exp\left(is\epsilon(t'-t'')-is\psi(t'^2-t''^2)\right)
\nonumber\\&\times&
\langle\delta\alpha(t')e^{isgt'}|\delta\alpha(t'')e^{isgt''}\rangle,
\end{eqnarray} 
and $\langle\alpha|\beta\rangle$ is the
overlap between coherent states, and
\begin{eqnarray}
\xi^{(2)}_s(t',t'')&=&\delta x^{(1)}_s(t')^*+\delta x^{(1)}_s(t''),\\
\zeta^{(2)}_s(t',t'')&=&-\delta x^{(1)}_s(t')+\delta x^{(1)}_s(t'').
\end{eqnarray} 
The first term $\xi^{(2)}_s(t',t'')$ represents the complex displacement 
of the oscillator position due to the perturbation acting one time at $t''<t$ 
(forward in time), and one time at $-t'>-t$ (backward in time). The second 
term $\zeta^{(2)}_s(t',t'')$ represents the displacement of the oscillator 
due to the perturbation acting two times at $t''<t'<t$. Between the two 
perturbations the system evolves freely for the time $t''-t'$ and 
accumulates a phase that depends on the difference of the effective 
qubit-dependent frequencies. In the short time approximation 
$t\approx 1/\epsilon$ such a phase can be neglected. Integrating the position 
degree of freedom over the subsets $\eta(s')$, we obtain
\begin{widetext}
\begin{equation}
F^{(2)}(s',t)_{ss}=-s's\frac{\epsilon^2}{4}
\int_0^tdt'\int_0^{t'}dt''
{\rm Re}\left\{{\cal O}_s(t',t'')
\left[{\rm erf}\left(\frac{\delta x(t)+\xi_{\bar s}^{(2)}(t',t'')}{\sigma}\right)
+{\rm erf}
\left(\frac{\delta x(t)+\zeta_{s}^{(2)}(t',t'')}{\sigma}\right)\right]\right\},
\end{equation}
where $\bar{s}=-s$. This expression has meaning only in the short time approximation. By setting $t\approx 1/\epsilon$, the correction $F^{(2)}(t)$ 
at second order to the discrete POVM is evaluated to be
\begin{equation}\label{Eq:F2}
F^{(2)}(t)=\frac{\epsilon^2}{4}
\int_0^tdt'\int_0^{t'}dt''
\cos\left(\epsilon(t'-t'')-\psi\left(t'^2-t''^2\right)\right)
e^{-\frac{1}{2}|A|^2(t'-t'')^2}
\left\{{\rm erf}[|A|(t+t'+t'')]+{\rm erf}[|A|(t-t'+t'')]\right\}.
\end{equation}
In an analogue way we calculate the elements of the first and second order 
contribution to the double measurement operator ${\bf C}$. The off-diagonal matrix element  of the first order 
contribution $C^{(1)}$ is
\begin{equation}
C^{(1)}(t';t)=\frac{1}{2}\left(\Gamma(t)-1\right)F^{(1)}(t'-t)
+\frac{i\Delta}{4}{\rm erf}\left(\frac{\delta x(t'-t)}{\sigma}\right)
\int_0^tdt'' e^{i\epsilon t''}\Gamma(t''),
\label{EqApp:C1}
\end{equation}
and the full expression of the diagonal matrix element of the second order 
contribution $C^{(2)}$ is
\begin{eqnarray}
C^{(2)}(t)&=&\frac{\epsilon^2}{8}
\int_0^tdt'\int_0^tdt''e^{i(\epsilon(t'-t'')-\psi(t'^2-t''^2))}e^{-\frac{|A|^2}{2}(t'-t'')^2}{\rm erf}[|A|(t+t'+t'')]\nonumber\\
&-&{\rm Im}\left[F^{(1)}(t)\epsilon\int_0^t
dt'e^{-i\epsilon t'}\Gamma(t)\Gamma(t')^*e^{-\frac{1}{2}|A t'|^2+|A|^2t't}
{\rm erf}\left(\frac{\delta x^{(1)}_-(t')}{\sigma}\right)\right].\label{Eq:C2app}
\end{eqnarray}
\end{widetext}


\begin{thebibliography} {99}
\bibitem{Note:coher-state-damp}
The factor 1/2 multiplying the damping rate $\kappa$ is typical of coherent
states arising from driving of a quantum harmonic oscillator,
subjected to Markovian damping due to weak coupling to a bath of
harmonic oscillator. The damping rate $\kappa$ can be related to the
quality factor $Q$, $\kappa=\omega_{\rm ho}/2\pi Q$ \cite{Scully}.
\bibitem{Note:erfB}
Strictly speaking, ${\rm erf}(x)$ approaches 1 only asymptotically, 
in the limit $x\rightarrow\infty$. However, for values of $x$ such that
 $1-{\rm erf }(x)<10^{-4}$, we discard the small deviation from 1.
\end{thebibliography}
\end{document}